\documentclass[12pt]{article}

\global\arraycolsep=1pt

\usepackage{color,amsbsy,amssymb,latexsym,amsfonts, amsmath,cancel}
\usepackage{braket}
\usepackage{mathrsfs}
\usepackage{graphicx}
\usepackage{cite}

\numberwithin{equation}{section}
\newcommand{\bel}[1]{\begin{equation}\label{#1}}                     
\newcommand{\bal}[1]{\begin{eqnarray}\label{#1}}   
\newcommand{\be}{\begin{equation}}               
\newcommand{\ba}{\begin{eqnarray}}           
\newcommand{\ee}{\end{equation}}
\newcommand{\ea}{\end{eqnarray}}

\renewcommand{\thefootnote}{\fnsymbol{footnote}}

\newcommand{\bea}{\begin{equation}}
\newcommand{\eea}{\end{equation}}

\begin{document}

\date{today}
%
%
\begin{titlepage}
\begin{flushright}
\normalsize
~~~~
OCU-PHYS 395\\
\end{flushright}

\vspace{15pt}

\begin{center}
{\Large\bf 
   126 GeV Higgs Boson 
   Associated with $D$-term Triggered \\ 
   \vspace*{2mm}
   Dynamical Supersymmetry Breaking }
\end{center}

\vspace{23pt}

\begin{center}
{\large H. Itoyama}$^{a, b} $
  and 
{\large Nobuhito Maru$^a$
}\\
%
\vspace{18pt}
%

$^a$ \it Department of Mathematics and Physics, Graduate School of Science\\
Osaka City University  and\\
\vspace{5pt}

$^b$ \it Osaka City University Advanced Mathematical Institute (OCAMI) \\

\vspace{5pt}

3-3-138, Sugimoto, Sumiyoshi-ku, Osaka, 558-8585, Japan \\

\end{center}
%
\vspace{20pt}
\begin{center}
{\bf Abstract} \\
\end{center}
%
 Continuing with our previous work on $D$-term triggered dynamical supersymmetry breaking \cite{IMaru1, IMaru2, IMaru3}, 
 we consider a system in which our generic ${\cal N}=1$ action is minimally extended 
 to include the pair of Higgs doublet superfields charged under the overall $U(1)$ 
 as well as $\mu$ and $B\mu$ terms. 
The gauge group is taken to be $SU(3)_C \times SU(2)_L \times U(1)_Y \times U(1)$. 
 We point out, among other things, that the Higgs mass less than the Z-boson mass at tree level 
 can be pushed up to be around 126 GeV by $D$-term contributions of the overall $U(1)$. 
 This is readily realized by taking a $U(1)$ gauge coupling of ${\cal O}(1)$.


\vfill

\setcounter{footnote}{0}
\renewcommand{\thefootnote}{\arabic{footnote}}

\end{titlepage}

\date{today}
\renewcommand{\thefootnote}{\arabic{footnote}}
\setcounter{footnote}{0}

\section{Introduction}


ATLAS \cite{ATLAS} and CMS \cite{CMS} collaborations recently announced 
 that a Higgs boson was discovered at the Large Hadron Collider (LHC) 
 and its mass is found to be around 126 GeV. 
Although the observed data for a variety of Higgs boson 
 decay modes are found to be consistent with the Standard Model (SM) expectations, 
 the new physics beyond the SM is indispensable for explanations of various unsolved issues in the SM 
 such as the origin of dark matter and dark energy. 
 
Supersymmetry (SUSY) is one of the leading promising candidates of new physics beyond the SM. 
In the minimal SUSY standard model (MSSM), 
 it is well known that the lightest Higgs boson mass at tree level is smaller than the Z-boson mass, 
 and can be enhanced up to 130 GeV through the quantum corrections by top and stop \cite{MSSMHiggs}.
The observed Higgs boson mass around 126 GeV requires a heavy stop mass or a large A-term for stop, 
 which leads to some amount of fine-tuning of parameters, {\it i.e.} ``little hierarchy problem".  
There are two typical extensions of the MSSM to overcome this issue. 
One is the next-to-MSSM (NMSSM) \cite{NMSSM} 
 and the other is the $U(1)$ extension of the SM gauge group \cite{ExU1}. 
In the former case, a gauge singlet chiral multiplet is introduced 
 and coupled to the Higgs doublets in the superpotential. 
$F$-term contributions can enhance the Higgs mass after developing the vacuum expectation value (VEV) of the singlet. 
In the latter case, $D$-term contributions can enhance the Higgs mass 
 if the Higgs doublets are charged under the extended $U(1)$ gauge group. 
We adopt here the latter approach since it can be naturally incorporated in our recently proposed mechanism 
 of $D$-term triggered dynamical SUSY breaking (DDSB) \cite{IMaru1, IMaru2, IMaru3}
 where a nonvanishing $D$-term VEV of the overall $U(1)$ gauge group is obtained 
 in Hartree-Fock analysis as is in the BCS model and the NJL theory \cite{BCS, NJL}.   
Also, our mechanism is a generalization of Dirac gaugino scenario \cite{Fayet, PS, HR, FNW}
 which has recently been paid attention to a solution to the natural SUSY breaking spectrum with 126 GeV Higgs mass 
 and many piece of work in various viewpoints have been done so far 
 \cite{NRSU, CFM, CFK, ABDQT, NPT, ABDQ, HNNS, Hsieh, KPW, BN, KMR, ABFP, BG, Blechman, 
 Carpenter, KOR, AG, DMRMC, BGM, FG, KM, Goodsell, Unwin, BGS, MNS, AB, DGHT}.  

In this letter, we investigate implications of the mechanism of DDSB uncovered in \cite{IMaru1, IMaru2, IMaru3}, 
 coupling the system to the MSSM Higgs sector which includes the $\mu$ and $B\mu$ terms.
The pair of Higgs doublet superfields $H_u, H_d$ is taken to be charged under the overall $U(1)$:
\ba
{\cal L}_{{\rm Higgs}} &=& \int d^4 \theta 
 \left[
 H_u^\dag e^{-g_Y V_1 - g_2 V_2 - 2e_u V_0} H_u 
 +H_d^\dag e^{g_Y V_1 - g_2 V_2 - 2e_d V_0} H_d 
 \right]  \nonumber \\
&& + \left[ \left(  \int d^2 \theta \mu H_u \cdot H_d \right)  
- B\mu H_u \cdot H_d + {\rm h.c.} \right] \;.
\label{HiggsLagrangian}
\ea
We have adopted notation $X \cdot Y \equiv \epsilon_{AB} X^A Y^B = X^A Y_A = - Y \cdot X$,
$\epsilon_{12} = - \epsilon_{21} = \epsilon^{21} = - \epsilon^{12} =1$. 
$V_{1,2,0}$ are vector superfields of the SM gauge group and that of the overall $U(1)$ respectively 
 and the corresponding gauge couplings are denoted by $g_{Y,2}$ and $e_{u,d}$ respectively.  
Unlike the MSSM case, the soft scalar Higgs masses $m_{H_u}^2 |H_u|^2, m_{H_d}^2 |H_d|^2$ are not introduced 
 since they are induced by $D$-term contributions in our framework.  

\section{Mechanism of D-term triggered dynamical supersymmetry breaking}
 Before going to the analysis on the objective of this letter, we summamrize here the basic qualitative features
 of the mechanism of D-term triggered dynamical supersymmetry breaking proposed in \cite{IMaru1, IMaru2, IMaru3}.
 
 Our underlying theory before the Higgs sector is coupled is given below by  eq. (\ref{DDSB}).  
By matching the tree part against the one-loop part in the effective potential (which is the Hartree-Fock approximation) 
 upon extremization with respect to the order parameter $\langle D^0 \rangle$, 
 we obtain the gap equation (eq. (21) of  \cite{IMaru1}, eqs.(4.26), (4.28) of \cite{IMaru3}). 
The gap equation is a self-consistency condition of the Hartree-Fock approximation and
  finding the explicit numerical solutions to the gap equation in \cite{IMaru1, IMaru3} demonstrates the self-consistency of the framework. 
Once $D$ term vev is generated, eq. of motion for the $D$ term tells that 
  the non-vanishing $D$ term vev implies the formation of the Dirac condensate,
 the reasoning of which is in parallel to NJL theory in the auxiliary field formalism. 
See eq.(2.10) of \cite{IMaru3}, and eq. below to eq.(4) of \cite{IMaru1}.

There are two fundamental scales in our original theory.  
The one is set by the mass parameter $M_{prep}$ contained in the prepotential function ${\cal F}$. 
The other is the mass parameter $M_{sup}$ contained in the superpotential $W$. 
The susy breaking scale, namely, the order parameter $\langle D^0 \rangle$ is found to be given by 
THEIR GEOMETRIC MEAN. 
$\langle D^0 \rangle \sim M_{prep} M_{sup}$. (See eq.(3.13) of \cite{IMaru3} for the derivation). 
So susy breaking scale can be arbitrarily large, depending upon how large these two parameters are. 
All of the adjoint multiplets of the standard model group appearing in our theory receive mass of order $M_{sup}$.  
(In this paper, we will not really consider including the MSSM matter sector 
belonging to the fundamental and anti-fundamental representations of the SM group: we only include the MSSM Higgs sector.) 

While the Hartree-Fock approximation exploits the one-loop effective potential in the auxiliary field formalism, 
the one-loop effective potential is matched to be in the same order as the tree level potential and this leads to the gap equation. 
The solution is transcendental in Planck constant and in this sense is nonperturbative. 
(It is well-known in the NJL type models that once the auxiliary fields are eliminated, 
this approximation is equivalent to the bubble summation of the fermion loops. 
See, for instance, \cite{GN}.)
Since we ignore the two-loop and higher in the auxiliary field formalism, 
quantum fluctuations are still assumed to be small on this new vacuum.
The solution to the gap equation itself, 
(which was obtained from the $D$ variation of tree and one-loop effective action), demonstrates the generation of 
susy breaking term. 
All these discussions are, of course, consistent with perturbative SUSY nonrenormalization theorem which 
applies to the $F$ term alone.

As we have already mentioned, there are two scales in our theory.
By making  $m \ll M_{prep} = {\rm cutoff~scale}$, we obtain $\langle D^0\rangle  \ll  ({\rm cutoff~scale})^2 $. 
Our gap parameter  $\Delta$ is a dimensionless parameter obtained from $\langle D^0\rangle$ 
and is determined to be ${\cal O}(1)$ by the gap equation. 
 We work, therefore, consistently in the weak field regime where our effective description is valid. 
Susy breaking scale is much larger than the electroweak scale but still much smaller than the cutoff. 


\section{Lagrangian and effective potential extended}
\subsection{Lagrangian}
Continuing \cite{IMaru3}, we work with
   the general ${\cal N}=1$ supersymmetric action consisting of chiral superfield $\Phi^a$ 
   in the adjoint representation and the vector superfield $V^a$
   that has been shown to break supersymmetry dynamically by the nonvanishing $D^0$ term :
  \ba
    {\cal L}_{{\rm DDSB}}
     &=&     
           \int d^4 \theta K(\Phi^a, \bar{\Phi}^a) + (gauging)   
        + \int d^2 \theta
           {\rm Im} \frac{1}{2} 
           \tau_{ab}(\Phi^a)
           {\cal W}^{\alpha a} {\cal W}^b_{\alpha}
            + \left(\int d^2 \theta W(\Phi^a)
         + c.c. \right).   \nonumber \\    
           \label{DDSB}
    \ea  
There are three input functions: 
 the K\"{a}hler potential $K(\Phi^a, \bar{\Phi}^a)$ with its gauging, 
 the gauge kinetic superfields $\tau_{ab}(\Phi^a)$ 
 that are the second derivatives of a holomorphic function ${\cal F}(\Phi^a)$,
 and a superpotential $W(\Phi^a)$.

As in \cite{IMaru3}, we make the following assumptions: 
\begin{itemize}
 \item[1)] third derivatives of ${\cal F}(\Phi^a)$ at the scalar VEV's are non-vanishing. 
 \item[2)] the superpotential at tree level preserves ${\cal N}=1$ supersymmetry.  
 \item[3)] the vacuum is taken to be in the unbroken phase of the gauge group 
$SU(3)_C \times SU(2)_L \times U(1)_Y \times U(1)$. 
\end{itemize}
The last assumption has been made for a technical reason and is not essential to the mechanism of 
dynamical supersymmetry breaking. 
   
   Some comments are in order for phenomenological applications of the theory based on this action.
First of all, the term we have proposed $\tau_{0a} {\cal W}^0 {\cal W}^a$ in our starting action eq.(\ref{DDSB})
 is very similar to the supersoft operator $A {\cal W}^{\prime} {\cal W}^a$ which appears in the more phenomenological operator analysis.
(See, for instance, \cite{FNW}.) 
The presence of the operator of this type alone has been known phenomenologically dangerous 
 as it leads to a massless particle in the imaginary part of the complex adjoint scalars.
The point we make here is, however, that our starting action eq.(\ref{DDSB})  consists not only of such part which contains this operator, 
 but also of the superpotential part which cannot be obtained just by an operator analysis. 
 The superpotential contains the scale $M_{sup}$ different both from the cutoff scale $M_{prep}$ 
 and from the electroweak scale contained in the Higgs sector and the resulting potential has a positive curvature everywhere near the extremum. 
 The ordinary $U(1)$ invariance of the complex scalar field is kept intact 
 and unbroken ${\cal N}=1$ supersymmetry of this term ensures that the tree spectrum obtained here is massive ${\cal N}=1$ supermultiplet 
 consisting of two real spinless particles and two polarization states of spin 1/2 particles.  
See, for instance, \cite{FIS3} for illustrative explicit computation of mass spectrum in the model of \cite{FIS1}.
So, All of the adjoint scalars in the standard model group receive masses of order $M_{sup}$ and the $D$ term supersymmetry breaking mechanism gives a boson-fermion splitting. 
There is no such light scalar in our theory to begin with in contrast to the operator analysis of \cite{FNW}.

It is also known that the operator of the type $\tau^{ab}\tau_{ab} {\cal W}^0 {\cal W}^0$ causes negative mass squared to the imaginary part of
 the adjoint scalars.
It is clear that our action does not contain such dangerous operator and our theory is free from such difficulty.


To simplify the analysis in what follows while keeping the essence, 
 we adopt the simplest prepotential and superpotential exploited 
 in \cite{IMaru3} of $5 \times 5$ complex matrix scalar superfield  $\varphi$ :
 \ba
{\cal F} =\frac{c}{2N}  {\rm tr} \varphi^2 + \frac{1}{3!MN} {\rm tr} \varphi^3\;, \;\;
W = \frac{m^2}{N} {\rm tr} \varphi + \frac{d}{3!N} {\rm tr} \varphi^3\;,
\label{f&W}
\ea
where $c$ is a pure imaginary number (as discussed in \cite{IMaru3}), and $m,M$ are mass parameters. 
Here $N=5$ and $M$ (real number) sets the scale in the prepotential, which is the cutoff scale.
    
We embed the generators of the gauge group into the bases which expand
  $\varphi$:
  \ba
 \varphi \equiv
\left(
\begin{array}{cc}
T_8 & 0 \\
0 &  T_3\\
\end{array}
\right)
   +  \sqrt{\frac{3}{5}} Y 
\left(
\begin{array}{cc}
 -\frac{1}{3} {\bf 1}_3 & 0 \\
0 &   \frac{1}{2} {\bf 1}_2   \\
\end{array}
\right)
  + \frac{ {\bf 1}_5}{\sqrt{10}} S \;, \;\;
T_3 = \sum_{a=1}^{3} T^a \left( \frac{\sigma^a}{2} \right)\;\;.
\label{varphiexpansion}
\ea 
  We have represented  the overall $U(1)$ and $U(1)_Y$  generators to be proportional to the unit matrix 
    and  the traceless diagonal generator respectively. 
  We analyze the case in which  only $S$  receives its VEV,
     namely, the unbroken $U(5)$ vacuum of the superpotential. 
We will make a comment for those cases in which these do not hold,
   which lead to the kinetic mixing.
We drop octet $T_8$ as it is irrelevant to the analysis below. 
    
After a simple calculation, we obtain  the non-vanishing prepotential derivatives
 \ba
 {\cal F}_{aa} &=&  \frac{c}{10} + \frac{3}{3! 5 \sqrt{10} M} \left( \sqrt{\frac{3}{2}} Y + S \right), \;
  {\cal F}_{00} = \frac{c}{10} + \frac{3 S}{3! 5 \sqrt{10} M}, \;  \nonumber \\
    {\cal F}_{YY} &=& \frac{c}{10} + \frac{3}{3! 5 \sqrt{10} M} \left( \sqrt{\frac{1}{6}} Y + S \right), \; 
    {\cal F}_{a0} =  \frac{3}{3! 5 \sqrt{10} M}  T^a, \;  \nonumber \\
     {\cal F}_{aY} &=&  \frac{3}{3! 5 \sqrt{10} M} \sqrt{\frac{3}{2}}
      T^a, \;   {\cal F}_{0Y} =  \frac{3}{3! 5 \sqrt{10} M} Y\;,
 \label{prep2nd}
 \ea
 their VEV's
 \ba
 \langle {\cal F}_{aa} \rangle  &=&  \langle {\cal F}_{YY} \rangle = \langle {\cal F}_{00} \rangle  =
  \frac{c}{10} + \frac{3}{3! 5 \sqrt{10} M}  \langle S \rangle , \; \nonumber \\
  \langle {\cal F}_{a0} \rangle  &=&  \langle {\cal F}_{aY} \rangle = \langle {\cal F}_{0Y} \rangle =0\;,
 \label{prep2ndvev}
 \ea
 and the derivatives of the superpotential
\ba
 \partial_a W &=&  \frac{3 d}{3! 5 \sqrt{10}} T^a \left( \sqrt{\frac{3}{2}} Y + S \right), \; \nonumber \\
 \partial_0 W &=&   \frac{m^2}{\sqrt{10}} +
  \frac{3 d}{3! 10 \sqrt{10}}  \left(  \sum_{a} T^aT^a +  Y^2 + S^2 \right), \; \nonumber \\
 \partial_Y W &=&  \frac{3 d}{3! 5 \sqrt{10}}  \left( \frac{3}{4} \sum_{a} T^aT^a +
 \frac{1}{4} Y^2 + \sqrt{\frac{3}{2}} S Y \right)\;.
\label{derisup}
\ea
 We choose $c= 10i$ but $\langle S \rangle $ is complex, not necessarily real.

In this letter, we add Eq. (\ref{HiggsLagrangian}) to Eq. (\ref{DDSB})
 and consider a part of ${\cal L}_{{\rm DDSB}} + {\cal L}_{{\rm Higgs}}$ relevant to 126 GeV Higgs
\ba
{\cal L} &=&  {\cal L}_{{\rm Higgs}} 
 + \int d^2 \theta
           {\rm Im} \frac{1}{2} 
           {\cal F}_{ab}(\Phi^a)
           {\cal W}^{\alpha a} {\cal W}^b_{\alpha}  \nonumber \\
&=& {\cal L}_{{\rm Higgs}} 
+ \frac{1}{4} 
\left[ 
\int d^2 \theta ( {\cal W}^a {\cal W}^a + {\cal W}^Y {\cal W}^Y + {\cal W}^0 {\cal W}^0 ) 
 + {\rm h.c.} \right] \nonumber \\
&&+\frac{1}{4} \left[ 
\int d^2\theta (
{\cal F}_{aaY} Y {\cal W}^a {\cal W}^a 
+ {\cal F}_{aa0} S {\cal W}^a {\cal W}^a 
+ {\cal F}_{YYY} Y {\cal W}^Y {\cal W}^Y 
+ {\cal F}_{YY0} S {\cal W}^Y {\cal W}^Y 
\right. \nonumber \\
&& \left. 
+ {\cal F}_{000} S {\cal W}^0 {\cal W}^0 
+ {\cal F}_{a0a} T^a {\cal W}^a {\cal W}^0 
+ {\cal F}_{aYa} T^a {\cal W}^a {\cal W}^Y 
+ {\cal F}_{0YY} Y {\cal W}^0 {\cal W}^Y 
)
+{\rm h.c.}
\right]. 
\label{ConcreteL}
\ea
The third prepotential derivatives, which are now real numbers, 
can be read off from Eq. (\ref{prep2ndvev}). 

In our analysis, we take that the value of $D^0$ VEV is determined essentially 
 by our Hartree-Fock approximation in \cite{IMaru3}. 
This source of supersymmetry breaking is then fed to the Higgs sector 
 and its effects are given by a tree level analysis. 
We will argue the validity of this procedure below.    

Let us make comment that the $B \mu$ term in eq.(\ref{HiggsLagrangian}) is generated 
 once we include an operator $({\cal W}^0/M_{prep})^2 H_u H_d$ 
 in the superpotential as an interaction term of our starting action. 
Here $M_{prep}$ is the scale we have introduced in the prepotential function ${\cal F}$ and is regarded as a cutoff scale. 
It is easy to see that $(D/M_{prep})^2 H_u H_d$ term is generated after $d^2 \theta$ grassmann integrations.  
Picking up the vev of $D$, 
 we conclude that $m^2 H_u H_d$ with $m \sim \langle D^0 \rangle/M_{prep}$ is generated in the potential. 

\subsection{Higgs potential and variations}
Let us extract the part relevant to the Higgs potential in (\ref{ConcreteL}). 
\ba
{\cal L}_{{\rm pot}} &=& 
|F_{H_u}|^2 + \left(- \frac{g_Y}{2} D^Y -e_u D^0 \right) |H_u|^2 -g_2 H_u^\dag \sum_a D^a \frac{\sigma^a}{2} H_u 
\nonumber \\
&& 
+ |F_{H_d}|^2 + \left( \frac{g_Y}{2} D^Y -e_d D^0 \right) |H_d|^2 -g_2 H_d^\dag \sum_a D^a \frac{\sigma^a}{2} H_d
\nonumber \\
&& 
- \left( \mu H_u \cdot F_{H_u} + \mu F_{H_d} \cdot H_d +B\mu H_u \cdot H_d + {\rm h.c.} \right) 
+ \frac{1}{2} \left( \sum_a D^a D^a + (D^Y)^2 + (D^0)^2 \right) 
\nonumber \\
&& 
+ \frac{1}{2} \sum_{A,B,C=a,Y,0} {\rm Im} ({\cal F}_{ABC} \varphi^C) D^A D^B 
+ \Gamma^{{\rm 1-loop}}(D^0)
\label{Dpot}
\ea
where $\varphi^C=(T^a, Y, S)$. 
The one-loop part of the effective potential in \cite{IMaru1, IMaru3} is denoted by $\Gamma^{{\rm 1-loop}}(D^0)$. 
Fermionic backgrounds are not needed in the potential analysis of Higgs and are not included in Eq. (\ref{Dpot}). 

Let us vary ${\cal L}_{{\rm pot}}$
with respect to the auxiliary fields, 
 replacing $\varphi^C$ by their VEV $\langle \varphi^C \rangle =(0, 0, \langle S \rangle)$. 
\ba
\delta D^a: && 0 = (1+{\rm Im}{\cal F}''' \langle S \rangle)D^a 
-g_2 H_u^\dag \frac{\sigma^a}{2} H_u -g_2 H_d^\dag \frac{\sigma^a}{2} H_d, 
\label{Da} \\
\delta D^Y: && 0 = (1+{\rm Im}{\cal F}''' \langle S \rangle)D^Y 
- \frac{g_Y}{2} |H_u|^2  + \frac{g_Y}{2} |H_d|^2, 
\label{DY} \\
\delta D^0: && 0 = (1+{\rm Im}{\cal F}''' \langle S \rangle)D^0 
-e_u |H_u|^2 -e_d |H_d|^2 + \frac{\partial \Gamma^{{\rm 1-loop}}(D^0)}{\partial D^0}. 
\label{D0} 
\ea
Note that ${\cal F}_{aa0}={\cal F}_{YY0}={\cal F}_{000} \equiv {\cal F}'''$  and 
 that Eq. (\ref{D0}) with $e_u = e_d = 0$ is in fact the gap equation of \cite{IMaru1, IMaru3}. 
 Eliminating the auxiliary fields (approximately), we obtain Higgs potential
\ba
V_{{\rm Higgs}} &=& \frac{g_2^2}{2(1+{\rm Im}{\cal F}'''\langle S \rangle)} 
\left( H_u^\dag \frac{\sigma^a}{2} H_u + H_d^\dag \frac{\sigma^a}{2} H_d \right)^2 
+ \frac{g_Y^2}{8(1+{\rm Im}{\cal F}'''\langle S \rangle)} \left( |H_u|^2 - |H_d|^2 \right)^2 
\nonumber \\
&&
+ \frac{1}{2(1+{\rm Im}{\cal F}'''\langle S \rangle)} 
\left( e_u |H_u|^2 + e_d |H_d|^2 - \left. \frac{\partial \Gamma^{{\rm 1-loop}}(D^0)}{\partial D^0} \right|_{D^0=D^{0*}} \right)^2 
\nonumber \\
&&
+ |\mu|^2 (|H_u|^2 + |H_d|^2) + (B \mu H_u \cdot H_d + {\rm h.c.}). 
\ea 
Here we have denoted by $D^{0*}$ the solution to Eq.(\ref{D0}) the improved gap equation. 
The deviation $\delta D^{0*}$ of the value from $D^{0*}$ in \cite{IMaru3} is in fact small 
by the ratio of electroweak scale and SUSY breaking scale. 
Therefore, we approximate the solution to the improved gap equation 
 by the value of $D^{0*}$ in \cite{IMaru3} denoted as $\langle D^0 \rangle$. 
Taking into account the fact that ${\rm Im}{\cal F}'''\langle S \rangle \sim \langle S \rangle/M \ll 1$, 
 we neglect the term ${\rm Im}{\cal F}'''\langle S \rangle$ at the leading order. 
The resulting Higgs potential at the leading order is given by
\ba
V_{{\rm Higgs}} 
&\simeq& 
\frac{g_2^2}{2} \left( H_u^\dag \frac{\sigma^a}{2} H_u + H_d^\dag \frac{\sigma^a}{2} H_d \right)^2 
+ \frac{g_Y^2}{8} \left( |H_u|^2 - |H_d|^2 \right)^2 
\nonumber \\
&&
+ \frac{1}{2} \left( e_u |H_u|^2 + e_d |H_d|^2 - \langle D^0 \rangle \right)^2 
+ |\mu|^2 (|H_u|^2 + |H_d|^2) + (B \mu H_u \cdot H_d + {\rm h.c.})
\nonumber \\
&=& 
\frac{g_2^2 + g_Y^2}{8} 
\left[
|H_u^0|^2 - |H_d^0|^2
\right]^2
+ \frac{1}{2} 
\left(
e_u |H_u^0|^2 + e_d |H_d^0|^2 - \langle D^0 \rangle 
\right)^2 \nonumber \\
&&+ |\mu|^2 \left( |H_u^0|^2 + |H_d^0|^2\right) 
- \left( B \mu H_u^0 H_d^0 + {\rm h.c.} \right) 
\nonumber \\
&=&
\frac{g_2^2 + g_Y^2}{32} v^4 c_{2\beta}^2 
+ \frac{v^2}{2} 
\left[ 
\mu^2 - B \mu s_{2\beta}
\right]  
+ \frac{1}{8}  
\left(
( e_u s_\beta^2 + e_d c_\beta^2)v^2 - 2 \langle D^0 \rangle  
\right)^2 
\ea 
where we have restricted the potential to the CP-even neutral sector 
of Higgs doublets $H_u=(H_u^+, H_u^0)^T$, $H_d=(H_d^0, H_d^-)^T$ 
in the second line 
 since we are interested in the Higgs mass. 
In the last line,   
 the neutral components of Higgs fields are defined as
\ba
&&H_u^0 = \frac{1}{\sqrt{2}} \left[ s_\beta (v+h) + c_\beta H + i (c_\beta A - s_\beta G^0 ) \right], \\
&&H_d^0 = \frac{1}{\sqrt{2}} \left[ c_\beta (v+h) - s_\beta H + i (s_\beta A + c_\beta G^0 ) \right]
\ea
and we use the shorthand notations: 
\ba
s_\beta \equiv \sin \beta, \quad c_\beta \equiv \cos \beta, \quad t_\beta \equiv \tan \beta, \quad 
s_{2\beta} \equiv \sin 2\beta, \quad c_{2\beta} \equiv \cos 2 \beta. 
\ea
$G^0$ is the would-be Nambu-Goldstone boson eaten as the longitudinal component of $Z$-boson. 
The VEV of Higgs field is $v \simeq 246$~GeV and $\frac{g_Y^2+g_2^2}{4}v^2=M_Z^2$ in this convention.

\section{Estimate of the Higgs mass}
We are now ready to calculate Higgs mass. 
As in the MSSM, 
 the minimization of the scalar potential $\partial V_{{\rm Higgs}}/\partial v^2 = \partial V_{{\rm Higgs}}/\partial \beta = 0$ 
 allows us to express $\mu$ and $B\mu$ in terms of other parameters. 
\ba
\mu^2 + \frac{M_Z^2}{2} &=& 
\frac{1}{2c_{2\beta}} \left( (e_u s_\beta^2 + e_d c_\beta^2 )v^2 - 2 \langle D^0 \rangle \right) 
\left(
e_u s_\beta^2 - e_d c_\beta^2
\right), \\
M_A^2 \equiv \frac{2B \mu}{s_{2\beta}} 
&=& 2 \mu^2 
+ \frac{e_u+e_d}{2} \left( (e_u s_\beta^2 + e_d c_\beta^2 ) v^2 - 2 \langle D^0 \rangle \right) \nonumber \\
&=& -M_Z^2 + \frac{e_u-e_d}{2 c_{2\beta}} 
\left( (e_u s_\beta^2 + e_d c_\beta^2 ) v^2 - 2 \langle D^0 \rangle \right). 
\ea
It is straightforward to obtain the mass matrix for CP-even Higgs from the second derivative of the potential,  
\ba
{\cal M}^2 =
\left(
\begin{array}{cc}
m^2_{hh} & m^2_{hH} \\
m^2_{hH} & m^2_{HH} \\
\end{array}
\right)
\ea
where each component is given by
\ba
m_{hh}^2 &=& M_Z^2 c_{2\beta}^2 + v^2 \left( e_u s_\beta^2 + e_d c_\beta^2 \right)^2, \\
m^2_{HH} &=& M_A^2 + M_Z^2 s_{2\beta}^2 
+ v^2 s_{2\beta}^2 
\left(
\frac{e_u-e_d}{2} 
\right)^2,\\
m_{hH}^2 &=& -M_Z^2 s_{2\beta} c_{2\beta} 
+ v^2 s_{2\beta} 
\left( e_u s_\beta^2 + e_d c_\beta^2 \right)
\left( \frac{e_u-e_d}{2} \right).  
\ea
The eigenvalues of this mass matrix are found as
\ba
\frac{1}{2} \left[ m_{hh}^2 + m^2_{HH} \pm \sqrt{ (m_{hh}^2 - m^2_{HH})^2 + 4m_{hH}^4 } \right] 
\ea
and the lightest CP-even Higgs mass is 
\ba
m_{{\rm Higgs}}^2 &=& \frac{1}{2} \left[ m_{hh}^2 + m^2_{HH} - \sqrt{ (m_{hh}^2 - m^2_{HH})^2 + 4m_{hH}^4 } \right]. 
\ea 
In order for the $\mu$-term to be allowed in the superpotential, 
 we must have a condition $e_u+e_d=0$ which is also required from an anomaly cancellation condition for the overall $U(1)$. 
Then, the Higgs mass can be expressed as
\ba
m_{{\rm Higgs}}^2 &=& 
\frac{1}{2} \left[ 
M_Z^2 + M_A^2 + e_u^2 v^2  
-\sqrt{\left( M_A^2-M_Z^2 c_{4\beta} - c_{4\beta} e_u^2 v^2 \right)^2 
+ s_{4\beta}^2 \left( M_Z^2 + e_u^2 v^2 \right)^2}
\right] \nonumber \\
&=& 
\frac{1}{2} \left[ 
\tilde{M}_Z^2 + M_A^2 
-\sqrt{\left( 
\tilde{M}_Z^2 + M_A^2 
 \right)^2 
 -4 \tilde{M}_Z^2 M_A^2c_{2\beta}^2 
 }
\right] 
\label{Hmass}
\ea
where 
$\tilde{M}^2_Z \equiv M_Z^2  + e_u^2 v^2$. 
It is interesting to see the correspondence between our expression of Higgs mass (\ref{Hmass}) and that in the MSSM,
\ba
m_{{\rm MSSM~Higgs}}^2 &=& 
\frac{1}{2} \left[ 
M_Z^2 + M_A^2 
-\sqrt{\left( M_Z^2 + M_A^2 \right)^2 
 -4 M_Z^2 M_A^2c_{2\beta}^2 }
\right]. 
\ea 
As in the case of MSSM, the upper bound of Higgs mass can be obtained 
 by taking a decoupling limit $M_A^2 \to \infty$, 
\ba
m^2_{{\rm Higgs}} \to \tilde{M}_Z^2 c_{2\beta}^2. 
\ea
$\tilde{M}_Z$ can be large enough by taking ${\cal O}(1)$ charge $e_u$ 
\ba
\tilde{M}_Z \sim \sqrt{(90~{\rm GeV})^2 + \left(246~{\rm GeV} \right)^2} \sim 262~{\rm GeV}. 
\ea
Let us go back to the minimization conditions of Higgs potential with $e_u+e_d=0$,
\ba
\mu^2+\frac{M_Z^2}{2} &=& \frac{e_u}{2c_{2\beta}} \left( -c_{2\beta}e_u v^2 - 2 \langle D^0 \rangle \right), \\
M_A^2 &=& 2\mu^2 
 = -M_Z^2 - \frac{e_u}{c_{2\beta}} 
\left( e_u c_{2\beta}  v^2 - 2 \langle D^0 \rangle \right)
\ea
which leads to
\ba
M_Z^2 + M_A^2 = -\frac{e_u}{c_{2\beta}} \left( c_{2\beta} e_u v^2 + 2 \langle D^0 \rangle \right).
\label{minimization}
\ea
In order to satisfy this condition, 
 the dominant part in the right-hand side of (\ref{minimization}) 
 $e_u \langle D^0 \rangle/c_{2\beta}$ is required to be negative. 

Using these conditions, we can eliminate $M_A^2$ in Higgs mass (\ref{Hmass}). 
\ba
m^2_{{\rm Higgs}} &=& \frac{1}{2} 
\left[
-\frac{2e_u}{c_{2\beta}} \langle D^0 \rangle 
- \sqrt{\left(-\frac{2e_u}{c_{2\beta}} \langle D^0 \rangle \right)^2 
+ 8 c_{2\beta} e_u \tilde{M}_Z^2 \langle D^0 \rangle + 4c_{2\beta}^2 \tilde{M}^4_Z}
\right] 
\nonumber \\
&\simeq&
\tilde{M}^2_Z c_{2\beta}^2 \left( 1+\frac{c_{2\beta} \tilde{M}^2_Z}{2e_u \langle D^0 \rangle} s_{2\beta}^2 \right) 
\ea
where the approximation $\langle D^0 \rangle \gg \tilde{M}_Z^2$ is applied in the second line. 

A plot for 126 GeV Higgs mass as a function of $\cos 2\beta$ and $e_u$ is shown below. 
Here we have taken $e_u < 0$ and $\cos 2\beta>0$ to satisfy the condition $e_u \langle D^0 \rangle/c_{2\beta}<0$.
We can immediately see that 126 GeV Higgs mass is realized by ${\cal O}(1)$ charge $e_u$, 
 namely without fine-tuning of parameters. 
Also, we found that the result is insensitive to the values of $D$-term VEV. 
This fact is naturally expected from the non-decoupling nature of Higgs mass.

\begin{figure}[htbp]
 \begin{center}
  \includegraphics[width=70mm]{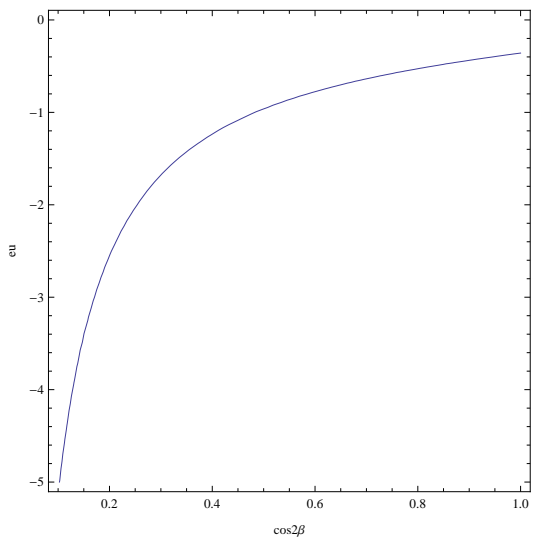}
 \end{center}
 \caption{A plot for 126 GeV Higgs mass as a function of $\cos 2\beta$ and $e_u$. 
 The result is insensitive to the values of D-term VEV.}
 \label{126GeV}
\end{figure}

\section{Summary}
In this letter, we have examined Higgs mass 
 in theory of $D$-term triggered dynamical SUSY breaking 
 minimally extended to couple the Higgs sector of the MSSM. 
Since the Higgs doublets are charged under the overall $U(1)$ in our framework, 
 the soft scalar Higgs masses are induced by the overall $U(1)$ $D$-term contributions 
 after SUSY breaking unlike the MSSM case. 
These $D$-term contributions can enhance the Higgs mass 
 which is less than $Z$-boson mass at tree level in the MSSM. 
We have shown that 126 GeV Higgs mass is naturally realized 
 by taking an overall $U(1)$ gauge coupling of ${\cal O}(1)$.    
 
We give a comment on anomaly cancellation of the overall $U(1)$ gauge symmetry
 which arises when the MSSM matter sector is added to our current construction. 
If we consider that the SM matter and Higgs are charged under the overall $U(1)$ 
 and the superpotential allows the SM Yukawa coupling and $\mu$-term, 
 then the overall $U(1)$ will be found to be anomalous. 
To resolve this problem, we need to introduce some additional fields charged under the overall $U(1)$. 
Actually, we can easily confirm that all of the anomalies for $U(1)(SU(3)_C)^2, U(1)(SU(2)_L)^2,$\\$U(1)(U(1)_Y)^2, U(1), (U(1))^3$ 
 are completely cancelled by introducing two SM singlets with ??? appropriate $U(1)$ charges, for instance.  
Yukawa coupling and $\mu$-term are allowed in the superpotential under the conditions
\ba
&& e_u + e_q + e_{\bar{u}}  = 0, \\
&& e_d + e_q + e_{\bar{d}}  = 0, \\
&& e_d + e_l + e_{\bar{d}}  = 0, \\
&& e_u + e_d = 0
\ea       
where $e_{q,l}$ is a $U(1)$ charge of the $SU(2)_L$ doublet quark, lepton superfields $Q, L$, 
 $e_{\bar{u}, \bar{d}}$ are those of the $SU(2)_L$ singlet quark superfields $\bar{U}, \bar{D}$, 
 and $e_{u,d}$ are those of Higgs doublet superfields $H_{u,d}$. 
The anomaly cancellation conditions are given  as follows. 
\ba
U(1)(SU(3)_C)^2:&& 2e_q + e_{\bar{u}} + e_{\bar{d}}  =0, 
\label{133} \\ 
U(1)(SU(2)_L)^2:&& 3e_q + e_l = 0 \to e_l = -3e_q, 
\label{122} \\
U(1)(U(1)_Y)^2:&& \frac{1}{36} e_q +\frac{4}{9}e_{\bar{u}} +\frac{1}{9}e_{\bar{d}} + \frac{1}{4} e_l + e_{\bar{e}} = 0, \\ 
\label{111}
U(1):&& 3 \left[ 6 e_q +3 (e_{\bar{u}} + e_{\bar{d}}) + 2 e_l + e_{\bar{e}} \right] + 2(e_u + e_d) + \sum_i q_i = 0, 
\label{grav1} \\
(U(1))^3:&& 3 \left[ 6e_q^3 +3(e_{\bar{u}}^3 + e_{\bar{d}}^3) + 2e_l^3 + e_{\bar{e}}^3 \right] + 2(e_u^3 + e_d^3) + \sum_i q_i^3 = 0.  
\label{gauge1}
\ea 
 where we have introduced additional singlet fields under the Standard Model gauge group and their $U(1)$ charge $q_i$. 
For instance, if we introduce two singlets with $U(1)$ charges $q_{1,2}$ 
 satisfying $q_1 q_2 = \frac{2}{3} \left( \frac{67}{31} \right)^2 e_u^2$, 
 we find all of the anomalies to be cancelled.  


\section*{Acknowledgments}
The work of H.I. is supported in part by the Grant-in-Aid 
 for Scientific Research from the Ministry of Education, 
 Science and Culture, Japan No. 23540316.
The work of N.M. is supported in part by the Grant-in-Aid 
 for Scientific Research from the Ministry of Education, 
 Science and Culture, Japan No. 24540283. 



\begin{thebibliography}{99}

\bibitem{ATLAS} 
  G.~Aad {\it et al.}  [ATLAS Collaboration],
  Phys.\ Lett.\ B {\bf 716}, 1 (2012). 

\bibitem{CMS} 
  S.~Chatrchyan {\it et al.}  [CMS Collaboration],
  Phys.\ Lett.\ B {\bf 716}, 30 (2012). 

\bibitem{MSSMHiggs}
Y.~Okada, M.~Yamaguchi and T.~Yanagida,
  Prog.\ Theor.\ Phys.\  {\bf 85}, 1 (1991);
  J.~R.~Ellis, G.~Ridolfi and F.~Zwirner,
  Phys.\ Lett.\ B {\bf 257}, 83 (1991); 
  H.~E.~Haber and R.~Hempfling,
  Phys.\ Rev.\ Lett.\  {\bf 66}, 1815 (1991).

\bibitem{NMSSM}
See for a review,  
  U.~Ellwanger, C.~Hugonie and A.~M.~Teixeira,
  Phys.\ Rept.\  {\bf 496}, 1 (2010). 
  
 \bibitem{ExU1}
See for a review,  
  P.~Langacker,
  Rev.\ Mod.\ Phys.\  {\bf 81}, 1199 (2009).
  
 \bibitem{IMaru1}
  H.~Itoyama and N.~Maru,
  Int.\ J.\ Mod.\ Phys.\ A {\bf 27}, 1250159 (2012). 
  
 \bibitem{IMaru2} 
  H.~Itoyama and N.~Maru,
  Int.\ J.\ Mod.\ Phys.\ Conf.\ Ser.\  {\bf 21}, 42 (2013) 
  [arXiv:1207.7152 [hep-ph]].

 \bibitem{IMaru3}
  H.~Itoyama and N.~Maru,
  Phys.\ Rev.\ D {\bf 88}, 025012 (2013). 
 
 \bibitem{BCS}
  J.~Bardeen, L.~N.~Cooper and J.~R.~Schrieffer,
  Phys.\ Rev.\  {\bf 108}, 1175 (1957);
  Y.~Nambu,
  Phys.\ Rev.\ Lett.\  {\bf 4} (1960) 380;
  Phys.\ Rev.\  {\bf 117}, 648 (1960).

\bibitem{NJL}
  Y.~Nambu and G.~Jona-Lasinio,
  Phys.\ Rev.\  {\bf 122} (1961) 345;
  %
  Phys.\ Rev.\  {\bf 124} (1961) 246.
 
\bibitem{Fayet} 
  P.~Fayet,
  Phys.\ Lett.\ B {\bf 78}, 417 (1978).
 
 \bibitem{PS} 
  J.~Polchinski and L.~Susskind,
  Phys.\ Rev.\ D {\bf 26}, 3661 (1982).
 
\bibitem{HR} 
  L.~J.~Hall and L.~Randall,
  Nucl.\ Phys.\ B {\bf 352}, 289 (1991). 

\bibitem{FNW} 
  P.~J.~Fox, A.~E.~Nelson and N.~Weiner,
  JHEP {\bf 0208}, 035 (2002). 
 
\bibitem{NRSU} 
  A.~E.~Nelson, N.~Rius, V.~Sanz and M.~Unsal,
  JHEP {\bf 0208}, 039 (2002). 

\bibitem{CFM} 
  Z.~Chacko, P.~J.~Fox and H.~Murayama,
  Nucl.\ Phys.\ B {\bf 706}, 53 (2005). 


\bibitem{CFK} 
  L.~M.~Carpenter, P.~J.~Fox and D.~E.~Kaplan,
  hep-ph/0503093. 
 
\bibitem{ABDQT} 
  I.~Antoniadis, A.~Delgado, K.~Benakli, M.~Quiros and M.~Tuckmantel,
  Phys.\ Lett.\ B {\bf 634}, 302 (2006). 
 
 
\bibitem{NPT} 
  Y.~Nomura, D.~Poland and B.~Tweedie,
  Nucl.\ Phys.\ B {\bf 745}, 29 (2006). 
 
\bibitem{ABDQ} 
  I.~Antoniadis, K.~Benakli, A.~Delgado and M.~Quiros,
  Adv.\ Stud.\ Theor.\ Phys.\  {\bf 2}, 645 (2008)
  [hep-ph/0610265]. 
 
\bibitem{HNNS} 
  J.~Hisano, M.~Nagai, T.~Naganawa and M.~Senami,
  Phys.\ Lett.\ B {\bf 644}, 256 (2007). 

\bibitem{Hsieh} 
  K.~Hsieh,
  Phys.\ Rev.\ D {\bf 77}, 015004 (2008).

\bibitem{KPW} 
  G.~D.~Kribs, E.~Poppitz and N.~Weiner,
  Phys.\ Rev.\ D {\bf 78}, 055010 (2008). 
 
\bibitem{BN} 
  A.~E.~Blechman and S.~-P.~Ng,
  JHEP {\bf 0806}, 043 (2008). 
 
\bibitem{KMR} 
  G.~D.~Kribs, A.~Martin and T.~S.~Roy,
  JHEP {\bf 0901}, 023 (2009). 
  
   
 \bibitem{ABFP} 
  S.~D.~L.~Amigo, A.~E.~Blechman, P.~J.~Fox and E.~Poppitz,
  JHEP {\bf 0901}, 018 (2009). 

\bibitem{BG} 
  K.~Benakli and M.~D.~Goodsell,
  Nucl.\ Phys.\ B {\bf 816}, 185 (2009). 


\bibitem{Blechman} 
  A.~E.~Blechman,
  Mod.\ Phys.\ Lett.\ A {\bf 24}, 633 (2009). 

\bibitem{Carpenter} 
  L.~M.~Carpenter,
  JHEP {\bf 1209}, 102 (2012). 

\bibitem{KOR} 
  G.~D.~Kribs, T.~Okui and T.~S.~Roy,
  Phys.\ Rev.\ D {\bf 82}, 115010 (2010). 

\bibitem{AG} 
  S.~Abel and M.~Goodsell,
  JHEP {\bf 1106}, 064 (2011). 

\bibitem{DMRMC} 
  R.~Davies, J.~March-Russell and M.~McCullough,
  JHEP {\bf 1104}, 108 (2011). 
  
\bibitem{BGM} 
  K.~Benakli, M.~D.~Goodsell and A.~-K.~Maier,
  Nucl.\ Phys.\ B {\bf 851}, 445 (2011). 

\bibitem{FG} 
  C.~Frugiuele and T.~Gregoire,
  Phys.\ Rev.\ D {\bf 85}, 015016 (2012)
 
\bibitem{KM} 
  G.~D.~Kribs and A.~Martin,
  Phys.\ Rev.\ D {\bf 85}, 115014 (2012). 

\bibitem{Goodsell} 
  M.~D.~Goodsell,
  JHEP {\bf 1301}, 066 (2013). 

  
\bibitem{Unwin} 
  J.~Unwin,
  Phys.\ Rev.\ D {\bf 86}, 095002 (2012).
  
\bibitem{BGS} 
  K.~Benakli, M.~D.~Goodsell and F.~Staub,
  JHEP {\bf 1306}, 073 (2013). 

\bibitem{MNS} 
  Y.~Morita, H.~Nakano and T.~Shimomura,
  PTEP {\bf 2013}, 053B02 (2013). 
   
\bibitem{AB} 
  S.~Abel and D.~Busbridge,
  JHEP {\bf 1311}, 098 (2013).
 
 \bibitem{DGHT} 
  E.~Dudas, M.~D.~Goodsell L.~Heurtier and P.~Tziveloglou,
  arXiv:1312.2011 [hep-ph].
 
\bibitem{GN}
 D.~J.~Gross and A.~Neveu,
  Phys.\ Rev.\ D {\bf 10}, 3235 (1974). 
   
\bibitem{FIS3}
K.~Fujiwara, H.~Itoyama and M.~Sakaguchi,
  Nucl.\ Phys.\ B {\bf 723}, 33 (2005). 
 
\bibitem{FIS1}
   K.~Fujiwara, H.~Itoyama and M.~Sakaguchi,
  Prog.\ Theor.\ Phys.\  {\bf 113}, 429 (2005). 
 
 \end{thebibliography}
\end{document}